\documentclass[12pt]{article}
\usepackage{graphicx}
\usepackage{url}
\begin{document}

\def\b{\bigskip}
\def\s{\smallskip} \def\c{\centerline}  \baselineskip=15 pt  \c {\bf Interference of macroscopic 
superpositions }\b \c {Italo Vecchi} 
\c {Vicolo del Leoncorno 5 - 44100 Ferrara -
Italy} \c { email: vecchi@isthar.com } \b 

{ \bf Abstract}: {\sl A simple experimental procedure based on the Elitzur-Vaidman scheme  is proposed  to test the persistence of macroscopic superpositions. It is conjectured that its implementation will reveal the persistence of superpositions of macroscopic objects in the absence of a direct act of observation.}\s
 
\s
\b

Schroedinger's cats have kept physicists mulling since their appearance in $ [1] $ and the
properties of macroscopic superpositions are still hotly debated. The alleged impossiblity to detect superpositions of macroscopic objects (e.g. cats which are at the same time dead and alive) has given rise in recent years to a theory of environment-induced decoherence ( see $ [7] $ for a detailed introduction to its different aspects and [9] for a survey of alternative approaches) which attributes the apparent absence of macroscopic superpositions  to environment-induced perturbations. We propose here a simple experimental procedure based on Elitzur-Vaidman's "interaction-free measurement" to test the persistence of macroscopic superpositions.  Our present experimental procedure  yields different outcomes in the case when state vector reduction takes place (e.g. the cat is either dead or alive) and in the case  where   macroscopic superpositions persist. We conjecture that the implementation of our experiment  will reveal the persistence of superpositions of macroscopic objects in the absence of a direct act of observation. Experiments on macroscopic quantum coherence have been proposed by other authors ( see $[14]$, $[15]$, $[9]$, $[13]$ ), but the argument  proposed here appears to be new and quite feasible.\s

Interaction-free measurement is considered in the Elitzur-Vaidman scheme ($ [4] $, see also $[6]$) and in earlier work by Dicke ($ [3] $) and Renninger ($ [2] $). In the Elitzur-Vaidman scheme, which has attracted considerable interest in recent years (see e.g. $ [8] $), a robust  interference technique is  used  to ascertain a system's unknown macroscopic properties.  The theoretical arguments motivating the present proposal are set forth in $ [5] $, but they are not strictly necessary for an understanding of this paper. It may suffice to recall that in $[5]$ the assumptions underlying current decoherence theory are examined and any "preferred" or "pointer" basis is shown to depend on the observer and not to be an intrinsic property of the physical system.\s  \begin {center} \includegraphics [scale=.8]  {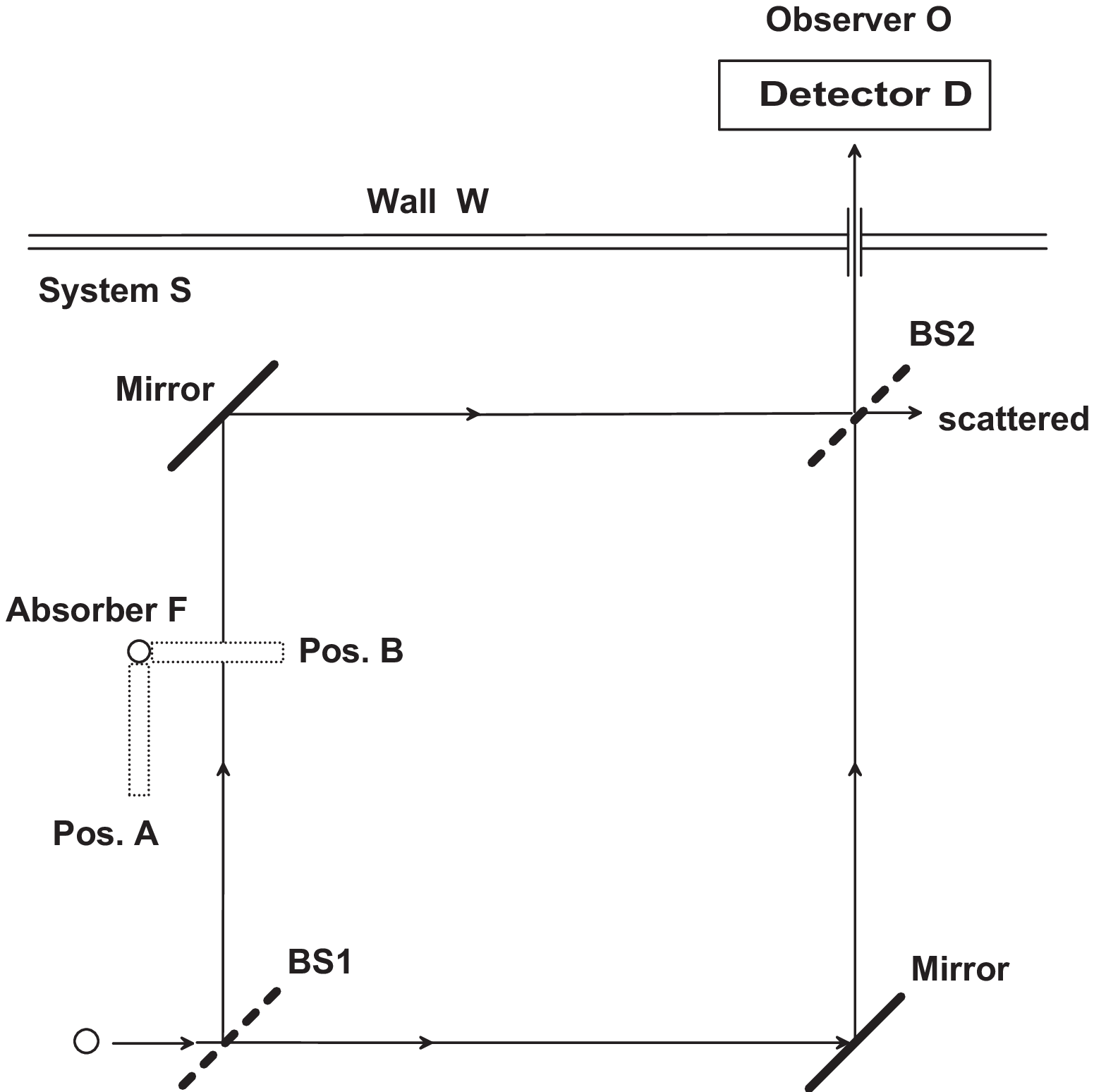} \end {center}
\c {\bf Fig. 1 } 

\s The  formalism of decoherence theory encodes
observer-dependent information into the system's evolution prior to measurement. The fact that
such encoding is  spurious is demonstrated by the procedure proposed here, which 
  implements a "tilted basis" measurement along the lines proposed in the final remarks of $ [5] $, so  that interference terms are not averaged out by random phase changes (cf. $[14]$, par. 2). In this setting the Elitzur-Vaidman scheme  reveals robust interference patterns that cannot  be detected in the most obvious "preferred basis". \s

Let us consider now the device described in Fig. 1.  The experimental setting, which is based on Mach-Zehnder interferometry, is  introduced in $ [4] $. 
 The small circle in the lower left corner is a photon source, emitting one photon at a time. The arrows refer to the photon's path while the thick hatched lines denote beam splitters, e.g. half-silvered mirrors. The dotted lines denote the possible positions of the absorber, as described below . \s

We adapt the procedure  of $ [4] $ by coupling the absorber  $ F $ to a quantum device $ QD $, which creates a superposition $\lambda_A |SA\rangle+ \lambda_B |SB\rangle $ of two states $ |SA\rangle$  and $ |SB\rangle $, corresponding to macroscopic events $ MEA $ and $ MEB $. In the original version of the Schroedinger's Cat experiment $ QD $ is a Geiger counter which is switched on for a given time and is connected to a relay that shatters a flask of hydrocyanidic acid. We may describe our quantum device $ QD $ as a Geiger counter which is switched on for a given time and is  connected to a relay that moves the absorber $ F $. In Schroedinger's original ([1]) argument  one has $ \| \lambda_A \|^2 = \| \lambda_B \|^2 = { 1 \over 2} $ and  the macroscopic events $ MEA $ and $ MEB $  are the shattering of cyanide flask or its remaining intact,  while $ |SA\rangle $ and $ |SB\rangle $ are the system's states where the Cat is respectively dead or alive. In our setting too $ \| \lambda_A \|^2 = \| \lambda_B \|^2 = { 1 \over 2} $, while  $ MEB $ corresponds to the insertion of the absorber $ F $ on the photon's path (Pos. B in Fig. 1) at time $ T=0 $, while $ MEA $ corresponds to the absorber $ F $ staying clear of the photon's path ( Pos. A  in Fig. 1). Accordingly  $ |SA\rangle$  corresponds to the absorber $ F $ being in position $ A $ while $ |SB\rangle$ corresponds to $ F $ being in position $ B$.\s

We assume that during each experimental run the system $ S $  beyond the wall $ W $ is isolated in such a way that the only information accessible  to the observer $ O $ is channelled through the detector $ D $. The system $ S $ does not need to be fully physically isolated from the observer $ O $ . Indeed the assumption that a system $ S $ be completely isolated is  impossible to realise in practice. What is needed in our setting however  is just the absence of a coherent communication channel between $ O $ and $ S $. In other words in our setting during the experiment the observer-experimenter has no way to know what is going on in system $ S $ behind the wall $ W $, excepts for the readings of the detector $ D $. \s

Both beam splitters $ BS1 $ and $ BS2 $  have transmission coefficient $ {1\over 2} $. The state of a photon moving to the right is denoted by $ |1\rangle $ and the state of a photon moving up is denoted by $ |2\rangle $. We denote as $ |absorbed\rangle $ the "empty wave" corresponding to photon's absorption.  Reflection induces a change of phase of $ {\pi\over 2}  $ the photon's wave function. The action of the beam splitter on the photon is ([4]) 
$$        
|1\rangle \longrightarrow { 1\over \sqrt 2}  ( |1\rangle + i |2\rangle), 
\qquad
|2\rangle \longrightarrow  { 1\over \sqrt 2} ( |2\rangle+ i |1\rangle)
$$
while the action of the mirrors is
$$ 
            |1\rangle  \longrightarrow i |2\rangle,
\qquad
            |2\rangle  \longrightarrow i |1\rangle.
$$

We want to verify whether at time $ T=1 $, when the photon is fired, the system subsists as a superposition of a a state $ |SA \rangle $ where the absorber  $ F $ is in position $ A $ and of a state $ |SB\rangle $ where the absorber in position $ B $. There are two  situations that we must consider. In the first situation the system  has already collapsed into one of the "macroscopically stable" states $ |SA\rangle $ and $ |SB\rangle $ so that one of the following two possibilities holds. \s

1a.  The absorber $ F $ is in position $ A $. The process is described by  $$ 
|1\rangle \longrightarrow  { 1\over { \sqrt 2} }( |1\rangle + i |2\rangle) \longrightarrow { 1\over \sqrt 2 } ( i|2\rangle - |1\rangle) $$ $$ \longrightarrow { 1\over 2} ( i|2\rangle - |1\rangle ) - { 1 \over 2} ( |1\rangle + i |2\rangle) = -|1\rangle
$$

so that the detector $ D $ never clicks  (cf. $[4]$). 

1b.  The absorber $ F$  is in position $ B $. The process is described by  
$$  |1\rangle \longrightarrow  { 1\over { \sqrt 2} }( |1\rangle + i |2\rangle) \longrightarrow { 1\over \sqrt 2 } ( i|2\rangle + |absorbed\rangle) $$ $$ \longrightarrow { 1\over 2} ( i|2\rangle -  |1\rangle ) + { 1\over \sqrt 2 } |absorbed\rangle $$ so that in this case  the detector $  D $ clicks with probability $ { 1 \over 4} $. 

Combining the above possiblities we see that, if the system $ S $ has already collapsed into one of the "macroscopically stable" states $ SA $ and $ SB $, then the probability that the  detector $  D $ clicks is $$ Pb(D)= { 1 \over 2 } {1 \over 4}=  {1 \over 8}. $$ This is the behaviour that we expect if  state vector reduction has already taken place when the photon is fired, i.e. if the system's state is a proper mixture of $ |SA\rangle $ (case 1a) and $ |SB\rangle $ (case 1b).

We consider now the case where the superposition of the states $ { 1\over \sqrt 2 } |SA\rangle $ and $ {1\over \sqrt 2} |SB\rangle $ corresponding to position $ A $ and position $ B $ subsists, but is perturbed by random interaction with the environment. As we will see, the measurement outcomes at the detector $ D $ are different from those arising when either case 1a or case 1b holds.

2. The absorber  $ F $ is in a superposition $ { a\over {\sqrt 2} }  |SA\rangle+ {b\over \sqrt 2} |SB\rangle$ of position $ A $ and position $ B $, where $ a $ and $ b $ are unknown uncontrollable phases such that $ | a |^2 = | b |^2 = 1 $ . In this situation the photon crossing the superposed absorber $ F $ undergoes a random phase change induced by the  the coupling between the photon's and the absorber's states.   The process is described by  $$ |1\rangle \longrightarrow  { 1\over { \sqrt 2} }( |1\rangle + i |2\rangle) \longrightarrow $$ $${ \alpha \over 2 \sqrt 2 } ( i|2\rangle - |1\rangle ) + { i\over \sqrt 2 } |absorbed\rangle - { 1 \over  2 }( |1\rangle + i |2\rangle) =  $$ $$ 
 { 1 \over 2 } \left [ \left ( { \alpha \over \sqrt 2 } -1 \right ) i |2\rangle  - 
\left ( { \alpha \over \sqrt 2 } + 1 \right ) i |1\rangle \right ]  + { i\over \sqrt 2 } |absorbed\rangle 
$$ 
where $ \alpha= \cos \theta + i \sin \theta $ is the photon's phase after it crosses the superposed absorber. Since  the phase change is random, $ \theta $ is spread uniformly on the interval $ [0, 2\pi]$ . Thus the probability that the detector $ D $ clicks is obtained  averaging on the $ [0, 2\pi]$ phase interval, i.e. on a randomly perturbed series of experimental runs $$ 
Pb(D)=
{ 1\over 2 \pi}  \int_0^{2\pi} \left |  { \cos \theta + i \sin \theta - \sqrt 2 \over 2 \sqrt 2 } \right|^2 \ d\theta= $$ $$
{ 1\over 16 \pi}  \int_0^{2\pi}   2 - 2 \sqrt 2 \cos \theta + \sin^2 \theta + \cos^2 \theta\ d\theta= { 3 \over  8 } 
$$     
so that the measurement outcomes at the detector  $ D $ are different from those where the system behaves as if state vector reduction had already taken place.   Heuristically we may describe the process in terms of "photon fractions", i.e. photon states with amplitude less than one, as follows.   One half of the half photon following the upper path is absorbed by  $ F $ and the other half reaches $ BS2 $ where its $ |2\rangle $ component combines with that of the half photon following the lower path. The standard (see $[14]$, par. 2) and plausible assumption is that the  phases of the photon's fractions which are not absorbed  are randomly perturbed as the photon becomes entangled with the absorber. The result is just an instance of Feynman`s rule that entangled particles do not exhibit interference other than with themselves ($[11]$, 1.6, III, see also $[17]$). In our case too "the effect of photons being scattered is enough to smear out any interference effect" ($[11]$, 1.6-11).  A random perturbation of the particle`s phase blurs any interference pattern, so that superpositions appear to vanish. It should be stressed however that the assumption that the upper photon´s phase is randomly parturbed as it crosses the absorber, while plausible and heuristically suggestive, is not an essential part of the argument. The key point here is that an entangled-perturbed particle does not exhibit interference with any other particle. It is the lack
of interference between the upper scattered
photon and the lower unperturbed photon that reveals the persistence of the absorber`s
superposition. \s

We  observe that the persistence of macroscopic superpositions yields different outcomes also under the assumption that the photon's phase is unaffected as it crosses the superposed absorber. Indeed, given a phase coefficient $ \alpha= \cos \theta + i \sin \theta $, the probablity that the photon be detected at $ D $ is 
$$
Pb(D)= {1 \over 8} | 3 - 2 \sqrt 2 \cos \theta | \geq { 3 - 2 \sqrt 2 \over 8 }
$$
where the equality holds only if $ \theta=0 $, i.e. if there is no coupling between the absorber's and the photon's phase. This case corresponds to a superposition of the solutions for  1a and 1b, but, since the initial conditions do not match those induced by the Geiger counter, the superposition principle is not applicable.
\s

Finally we observe that, when the information on the sytem $ S $ is extracted by the observer through the detector $ D $, the absorber's state no longer corresponds to a unitary evolution, so that thereafter the condition $ \| \lambda_A \|^2 = \| \lambda_B \|^2 = { 1 \over 2} $ may be violated. The system must be re-inizialized after each  measurement  and then the procedure can be applied again. As long as the unitarity
condition holds, i.e. as long as the device  described here can be imbedded in a
system whose evolution is unitary, the above argument holds. We refer to $[10]$, where a relevant argument pointing out the implications of the superposition principle at the macroscopic level is presented. \s

Our argument shows  that in this setting entanglement prevents the system from duplicating the experimental outcomes of state-vector reduction. We conjecture that the experimental outcomes will confirm the persistence of macroscopic superpositions,  as in case 2. A measured value $ Pb(D) \neq { 1\over 8} $  would confirm the persistence of superpositions. An experimental outcome of $  Pb(D)= { 1\over 8} $ would provide strong evidence that state vector reduction has already taken place when the photon crosses the absorber, although it would not directly imply that this is the case, since, as shown above, there are values of $ \theta $ which yield $ Pb(D) \leq { 1 \over 8}$, so that some peculiar distribution of phase changes corresponding to some coherent photon-absorber coupling might yield the same measurement outcome as state vector reduction.\s

It is worth stressing that, although the environmental states induced by the absorber are  orthogonal,  they do not correspond to the photon states which are being measured at the detector $ D $. The measurement does not take place in the "macroscopic" basis corresponding to the absorber's position, but in the basis relative to the photon state at  $ D $, whereas the photon state which is being measured is just the projection of the system's state vector on the photon's state space at $ D $.  A heuristic example may clarify the issue. Suppose there are boys and girls in a room. The boys  throw apples out of the window, the girls throw both apples and oranges. If the observer walks down the street and is hit by an orange he knows that it was a girl, if it is an apple he does not  know whether it was a boy or a girl. Boys and girls are orthogonal, but the observer is measuring them through a tilted basis. In this case case absorber states are boys vs. girls, whereas photon states at $ D $ are apples and oranges. \s
We stress that the slanted correspondence between absorber and photon states is induced by the lower unperturbed photon. Indeed we can modify the device arranging the absorber  so as to block the photon`s upper path when it it is in Pos. A and the lower path when it is in Pos. B.  Both the upper and the lower photon are scattered and become entangled and so they do not interfere. In this case the blurring of superposition does indeed mimic collapse, as in $[14]$.  \s

One may consider the possibility that the photon phase is affected by the environment in such a way as to mimic collapse. The experimental outcomes for the device in Fig. 1 corresponding to collapse can be duplicated only assuming that the superposed absorber  affects the lower photon's phase in a very specific way. One has to assume that the lower photon is split into two equal amplitudes, one of which is left unperturbed, while the other undergoes exactly the same random changes that affect the phase of the upper photon as it crosses the absorber. It is easy to see that in this case the experimental outcomes are the same as in the case where state-vector reduction takes place: the perturbed photons cancel out and the unperturbed one is just what is needed to mimic collapse.  Simply assuming that the superposed absorber induces a random perturbation of the lower photon's phase is not enough to mimic collapse, since it yields the same outcomes as in our conjecture, i.e. with the upper and lower photon not interfering. The superposed absorber action on the far-away lower photon must be both  strong and fine-tuned. The problem with this  model is that it attributes to superpositions quite extraordinary  physical properties, which distinguish them from "ordinary" states. Since the difference between a superposition and an "ordinary" (i.e. non-superposition) state depends only on the chosen basis this is quite perplexing, as the system would have to exhibit extraordinary physical properties in order to meet the observer's expectations. In reality particle's phases are quite sturdy and the can be changed only through strong interaction. It is precisely the relative insensitivity of particle's phases to environmental perturbations that makes Zehnder's interferometry possible.\s

It is worth pointing out that the argument presented in this letter appears related to  Maris`s recent work on  fractional electrons in liquid helium $[12]$, which has attracted considerable interest. The  phenomena examined in $[12]$  depend on a tilted-basis measurement, as superposed "electrino bubbles" are generated by delocalised electrons (cf. $[16]$).  The persistence of such bubbles, which carry a whole  electron charge , is revealed in the photoconductivity experiments by ionic mobility measurements. The ionic mobilty of superposed "electrino bubbles"  is increased since they are smaller than non-delocalised bubbles.  The second paragraph of $[12]$ deals with   questions  on quantum measurement theory that are related to those addressed here.
 The "tilted-basis" measurement described here is characterised, similarly to those described  in $[12]$ , by "non-vanishing dispersion", as pointed out in $[16]$. 
\b

 \c{ \bf References} \b

[1] E. Schroedinger, Naturwissenschaften 23, 807 (1935). \s

[2] M. Renninger, Z. Phys. 158, 417 (1960). \s

[3] R.H. Dicke, Am. J. Phys. 49, 925 (1981). \s

[4] A. Elitzur and L. Vaidman, Found. Phys. 23, 987 (1993). \s

[5] I.Vecchi, at \url{http://xxx.lanl.gov/abs/quant-ph/0002084} (see also: \url{http://xxx.lanl.gov/abs/quant-ph/0102130}). \s

[6] P.Kwiat, H. Weinfurter, T. Herzog and A. Zeilinger, Phys. Rev. Lett. 74(24), 4763 (1995). \s

[7] "Decoherence and the Appearance of a Classical World in Quantum Theory" D.Giulini et al. ed. , Spinger, Berlin-Heidelberg-New York 1999.\s

[8] R. Penrose "Shadows of the Mind", Oxford University Press, Oxford, 1994.\s

[9] G.Ghirardi, at \url{ http://xxx.lanl.gov/abs/quant-ph/9810028} (1998).\s

[10] A.Bassi and G.Ghirardi, at \url{ http://xxx.lanl.gov/abs/quant-ph/0009020} (2000).\s

[11] "The Feynman Lectures on Physics" , Addison-Wesley, 1963.\s

[12] H.J. Maris, Journal of Low Temperature Physics 120, 173.\s

[13] A.J. Leggett, Found. Phys. 29, 447 (1999) \s

[14] E.H. Walker et al. , Physica B, 339 (1988) \s

[15] F. Thaheld, Physics Letters A, 273, 232 (2000) \s

[16]  R. Jackiw, C. Rebbi and  J.R. Schrieffer, at \url {http://xxx.lanl.gov/abs/cond-mat/0012370} 

[17] A. Zeilinger, Rev. Mod. Phys. 71, No.2 (1999).

 \s

\end{document}